\documentclass[aps,prd,preprintnumbers,nofootinbib,twocolumn]{revtex4}
\usepackage{bm}
\usepackage{latexsym}
\usepackage{dcolumn}
\usepackage{amsmath,amsfonts,amssymb}
\usepackage{graphicx,epsfig}
\usepackage{psfrag}
\usepackage{color}
\usepackage{amsthm}

\def\be {\begin{equation}}
\def\ee {\end{equation}}
\def\bea {\begin{eqnarray}}
\def\eea {\end{eqnarray}}
\def\bc {\begin{center}}
\def\ec {\end{center}}
\def\bfg {\begin{figure}}
\def\efg {\end{figure}}
\def\bi {\begin{itemize}}
\def\ei {\end{itemize}}
\def\nn {\nonumber}

\def\le {\left}
\def\ri {\right}
\def\pa {\partial}

\def\no {\noindent}

\def\vs {\vspace}

\def\m  {\mu}
\def\n  {\nu}

\def\t  {\tau}

\def\beq{\begin{equation}}
\def\eeq{\end{equation}}
\def\br{\begin{eqnarray}}
\def\er{\end{eqnarray}}
\newcommand{\eel}[1] {\label{#1}\end{equation}}

\newcommand{\bdm}{\begin{displaymath}}
\newcommand{\edm}{\end{displaymath}}

\begin{document}
\title{The central role of symmetry in physics}

\author{Saurya Das} \email{saurya.das@uleth.ca}

\affiliation{Theoretical Physics Group, Department of Physics and Astronomy, and Quantum Alberta,
University of Lethbridge, 4401 University Drive,
Lethbridge, Alberta, Canada T1K 3M4}

\author{Gabor Kunstatter}\email{g.kunstatter@uwinnipeg.ca}

\affiliation{Winnipeg Institute for Theoretical Physics and
Department of Physics, The University of Winnipeg, 515 Portage Avenue, Winnipeg, Manitoba, Canada R3B 2E9}

\vspace{0.1cm}

\begin{abstract} %
\begin{center}
{\bf Invited review article for Journal of Applied and Fundamental Sciences \\
(Assam Don Bosco University, India)}

\end{center}

\vspace{0.1cm}
Spacetime and internal symmetries can be used to severely restrict the form of the equations for the fundamental laws of physics. The success of this approach in the context of general relativity and  particle physics motivates the conjecture that symmetries may help us to one day uncover the ultimate theory that provides a unique, unified description of all observed physical phenomena. We examine some of the strengths and weaknesses of this conjecture.

\end{abstract}

\maketitle

\section{The symmetric Universe}

Ever since Galileo announced to the world in the seventeenth century, at considerable risk to his life and liberty, that the earth does not occupy a special position in the then visible Universe,
the idea has been generalized much further.
For example, it was discovered that our solar system, with the Sun at its focus,
is but one of about $300$ billion such systems, drifting under the effect of the mean gravitational force that it experiences in the Milky Way galaxy.
The latter is part of a local supercluster, one of the tens of millions in our visible Universe. Furthermore, all the observed `clumpiness' in our Universe, in terms of stars, planets, galaxies etc., virtually disappears at scales of over $300$ Mpc or so. At larger scales the Universe appears as a smooth fluid described by a simple perfect fluid equation of state $p=w\rho$, where $p$ and $\rho$ are the local density and pressure in the so-called `co-moving frame' (in which points at rest follow geodesics), and $w$ is a constant.
There is now very strong evidence
that (i) our Universe is homogeneous and isotropic at large scales, and that its average $\rho$ almost exactly equals the critical density, i.e.
$\rho=\rho_{crit}= 3H^2/8\pi G= 10^{-26}~kg/m^3$ (where $H=$ Hubble parameter, $G=$ Newton's constant),
such that it is also spatially flat to a high degree of accuracy, and
(ii) about $0.7~\rho$ can be attributed to dark energy {(likely an inexplicably small cosmological constant $\Lambda$),}
$0.25~\rho$ to dark matter (there are many viable, but unproven candidates) and $0.05~\rho$ to visible matter, making up planets, stars and everything visible in between \cite{perlmutter}.
In other words, at cosmological length scales, the Universe looks the same everywhere
\footnote{{Most experimental evidence proves only isotropy. Homogeneity follows from the assumption of the Copernican principle, which states that we don't occupy a special place in the Universe. If it looks isotropic here on Earth, it must do so everywhere. Some direct experimental evidence for large-scale homogeneity does nonetheless exist \cite{Homogeneity}. }}, and in all directions, and there is no place in it, including our own, any more special than another \footnote{Cosmological models  that assume only spherical symmetry and not homogeneity do exist\cite{moffat}. These models necessarily violate the Copernican principle.}.


The high degree of symmetry at large scales makes the state of the  Universe very special and therefore requires explanation. Is it due to a miraculous fine tuning of initial conditions, or was it achieved dynamically from less special initial data as a natural consequence of physical laws? \footnote{The currently accepted explanation for this high degree of symmetry is the inflationary paradigm.  While it provides a fairly compelling possible resolution to several fundamental problems  and attempts have been made to compare with cosmological observations,
 it is not yet a theory {\it per se} because the precise mechanism for inflation is not yet known:
 the simplest models fail, while the successful ones generally have too much freedom to make unique predictions.
 \cite{steinhardt}.}

\section{The laws of physics must also be symmetric}

Our Universe provides the ultimate example of a physical system (matter and energy distribution) that possesses a remarkable degree of symmetry. The existence of such (approximate) symmetries is vital to our ability to study complex systems. By assuming, to a first approximation, that the symmetry is exact, it is possible to eliminate a large number of non-essential degrees of freedom, and thereby make a difficult problem tractable.

At a more fundamental level, the laws of physics themselves possess a high degree of symmetry.
At this point we must distinguish between symmetries of the state of a system, and symmetries of the physical laws describing that system. Symmetry of the laws do not require all physical states to possess the same symmetry. However, the lowest energy or ``ground" state is in general symmetric. If this is not the case, then the symmetry is said to be spontaneously broken.

One commonly cited symmetry that the laws of physics are thought to possess is coordinate invariance. In fact, this is not so much an assumption of symmetry but a pragmatic statement that we must be able to write the laws of physics in a form on which  all observers can agree, independent of the coordinates that they happen to use to describe the world\footnote{G.K. is grateful to J. Zanelli for enlightening conversations on this point.}.
In mathematical terms it says that the laws of physics can be written in tensor form and derived from a coordinate invariant action.

{A more physical and hence restrictive assumption is that there are no preferred observers, with associated coordinates, for whom the laws of physics take a fundamentally different form. In the context of a tensor formulation, this is equivalent to demanding that there be no background fields in nature that must be prescribed {\it a priori}. Although in the presence of such background fields the equations could still be written in tensor form, they would in general look very different in different coordinate systems. All the measurable fields in a viable physical theory must therefore transform as tensors and be dynamically determined by equations of motion from generic initial data. Without this assumption, the determination of the correct physical laws could in principle require an infinite number of measurements, at all times and places, in order to determine the values of the prescribed fields.
 We will henceforth call this assumption observer independence, to distinguish it from the less stringent assumption of coordinate invariance. }

Special relativity is a framework for constructing theories that assumes  observer independence for a special class of observers, namely those in inertial frames,  for whom the laws of physics simplify. Einstein's profound insight could be summarized as the realization that inertial observers will only agree on the laws of physics if the Lorentz transformations that are symmetries of Maxwell's equations  also leave all other laws invariant. Many expositions of special relativity therefore start with two fundamental assumptions:
\begin{enumerate}
\item The laws of physics are the same in all inertial frames
\item  The speed of light is a universal constant. This is justified by the observation that the speed of light in Maxwell's equations is determined by fundamental properties of the vacuum.
\end{enumerate}
However, as shown in a beautiful paper by Mermin
\cite{mermin1984}
 the velocity transformation laws of special relativity can be derived without any reference to Maxwell's equations or the velocity of light. This derivation starts from the single assumption that if  energy and momentum are conserved in one inertial frame then they must be  conserved in every frame. Since energy and momentum conservation follow via Noether's theorem (see below) from invariance of physical laws under temporal and spatial translations, Mermin's article proves that special relativity follows directly from the single assumption that the laws of physics are the same for every inertial observer. The two postulates normally cited as  fundamental to special relativity therefore reduce to this single fundamental assumption.

Extending the principle of observer independence to all observers, not just inertial, gives rise to the general theory of relativity. Special relativity contains a non-dynamical, prescribed field, that appears to violate the assumption of observer independence. This field is the Minkowski metric which determines the  geometry of space-time to be Lorentzian and flat.   Einstein's insight in the present case can be described as the realization that the principle of observer independence of the laws of physics requires the metric, i.e. the geometry of space-time itself, to be dynamical.

It is thought that the above considerations of spacetime symmetry must apply down to the smallest length scales,  including the electroweak or quark scale, and all the way down to  the Planck scale, $\ell_{Pl} \approx 10^{-35}~m$.
\footnote{The Planck scale is where quantum gravity is expected to become important. Below this scale it is possible that the notion of spacetime itself loses meaning, and with it the notions of spacetime symmetries.}
There is no reason to suspect that near these fundamental scales, any point in spacetime, any direction, or for that matter, any velocity would be preferred. It is nonetheless an open question as to what happens below the Planck scale. As discussed in Subsection \ref{section:quantum gravity},
this question must be addressed and answered by any viable theory of quantum gravity.

To summarize, symmetries of physical systems abound and are useful for solving complex problems. In addition, both coordinate invariance and observer independence of the laws themselves are necessary  for us to be able to formulate sensible theories. More importantly, these assumptions place useful restrictions on the form these theories can take. For example, as will be elaborated further on, the assumption of Lorentz invariance singles out Maxwell's equations, whereas more general observer independence leads uniquely to Einstein's equations in four space-time dimensions.

We now turn to the question of the extent to which one can use spacetime and other {\it internal} symmetries to limit the potentially infinite variety of theories at our disposal to just one or perhaps a single class. To paraphrase Einstein, what interests us is ``whether God had any choice in the creation of the World''.

\section{Order of differential equations}

The known fundamental laws of physics are
differential equations no higher than second order: Newton's second law - second order in time; Maxwell' equations and Yang-Mills theories - second order in space and time when written in terms of the potentials $A_\mu$,
Einstein equations - second order in space and time when written in terms of the metric, Friedmann
equation in cosmology -
second order in time,
Schr\"odinger equation - first order in time and second order in space; the Klein-Gordon equation - second order in space and time; the Dirac equation - first order in space and time. No fundamental law of nature appears to need greater than second order to express itself. While we are not aware of any fundamental reason, or a ``no-go" theorem as to why this is the case, one notes that higher orders would imply greater instabilities in the systems under consideration, which presumably nature does not prefer. After all, nature seems to be stable in the big scheme of things, and some degree of stability is again a necessary condition for us to be able to do physics. Although higher order theories abound in the physics literature,
they do not seem to be fundamental, and
we henceforth restrict consideration to those that are second order.

\section{Degrees of Freedom}

The fundamental physical laws describe interactions (forces), including self-interactions in most cases,  between a veritable zoo of observed (and in some instances, hypothesized) particles.
The strength of an interaction is characterized by the charge of the particle, or equivalently, a coupling constant, while the interactions themselves are mediated by fields. Quantum mechanics requires the charge carriers (particles) to be  represented by fields as well. These fields, taken together, constitute the fundamental degrees of freedom of the theory.

Going over the four known forces of nature, electromagnetism,
weak and strong nuclear forces are described by the so-called Yang-Mills, or non-abelian gauge theories. In these, forces are mediated by the Yang-Mills
vector fields $A_\mu \equiv A_\m^a\t_a$, where $\t_a$ are generators of the unitary gauge groups $U(1)$ ($a=1$), $SU(2)$ ($a=1,2,3$)
and $SU(3)$ ($a=1,\dots,8$) respectively. Some of these vector fields acquire masses, as we shall see
in the next section.

The fundamental matter degrees of freedom are described by the following fermionic fields, and their quanta:
$6$ leptons (electrons, muons, tau particle,  with mass about $0.5,105$ and $1800~MeV/c^2$ respectively, and electron neutrinos, muon neutrinos and taon neutrinos, a few $eV/c^2$ each),
$6$ quarks, named {\it up, down, strange, charm, bottom} and {\it top}, with masses about $2,4,101,1270~MeV/c^2$ and $4,172~GeV/c^2$ respectively,
and the complex Higgs bosonic scalar field of mass $125~GeV/c^2$.
Strong interactions also require that
each quark comes in three colours,
red, green, and blue.

The fourth fundamental force, namely gravity, is most commonly described by the second rank metric tensor $g_{\mu\nu}$, which  measures the distance between two spacetime points, or events.  The metric provides the generalization of Pythagoras' theorem to more interesting geometries than Euclidean. Alternative descriptions of gravity include the so-called tetrads, or vierbeins, which are roughly `square roots' of the metric, and defined as $e^a_\mu e_{a\nu} = g_{\mu\nu}$ and $e^{a\mu} e_{b\mu} = \delta^a_b$, where $a,b$ are indices on the flat tangent space, that can be constructed in every point in spacetime.

\section{Internal symmetries and gauge fields}
\label{section:gauge fields}

Non-relativistic quantum mechanics describes the state of  a single particle at a given time $t$  by a complex-valued wave function $\psi(\vec r;t)$
(relativistically, one uses a multi-component wavefunction, a column vector or `spinor', but much of the following arguments go through. The same is true for a multiparticle quantum system).
Since the probability density of observing the particle at a point $\vec{r}$ is given by $|\psi(\vec{r},t)|^2$,
the state of the particle is  unaffected by a change of phase: $\psi(\vec{r},t)\to e^{i\alpha} \psi(\vec{r};t)$, for any real constant $\alpha$. This is the first, and in many ways archetypal, example of an internal symmetry operation. It does not affect spacetime coordinates but instead changes the mathematical description of the state of the particle. Since $\alpha$ is a constant, the transformation acts simultaneously and identically at every point in space. It is  referred to as a global transformation.

Given the many particles that underlie the standard model it is natural to generalize the above transformation to one that not only changes the phase of each particle, but produces linear combinations of a set of $N$ wave functions $\psi^{(a)}$, $a=1..N$. The set of wave functions behaves as a vector in an internal space in which the symmetry transformations act as rotations. In order for probabilities to be preserved
the transformation must correspond to a group valued matrix $U \in SU(N)$ that takes $\psi^{(a)}\to \sum_{b=1}^N U^a_b\psi^{(b)}$, or in compact notation $\psi \to U\psi$. {In fact, $U$ can take its value in any symmetry group that has a unitary representation.}

So far we have been talking about  global transformations whose parameters are constant in space. One can ask whether these transformations can be made independently at each point in space, or spacetime. In the case of internal symmetries, one finds that this can and must in fact be done, but at the expense of introducing one gauge potential for each independent symmetry operation. Invoking the principle that there be no fixed, background fields, the gauge potentials must have dynamics of their own. Maxwell's equations can therefore be thought of as describing the dynamics of the gauge potential $A_\mu$ that emerges when the phase invariance of single particle quantum mechanics is made local, or ``gauged''\footnote{Remarkably in 1929 Weyl\cite{Weyl}  used general relativity and the existence of spinors to argue that the gauge potential (and hence electromagnetism) are a necessary consequence of local Lorentz invariance.  For excellent reviews of the history of gauge theories see \cite{Oraif, Jackson}}. The associated field is strength $F_{\mu\nu}= \partial_\mu A_\nu-\partial_\nu A_\mu$, with Lagrangian ${\mathcal L} = - \frac{1}{4} F_{\mu\nu}F^{\mu\nu}$. This is the only Lorentz and gauge invariant action in four dimensions that yields second order equations for $A_\mu$.

In four dimensions one can add another term can be added that neither violates  the proposed symmetries nor produces higher order equations:
\be
{\cal L}_\theta\propto \theta Tr(F_{\mu\nu}F_{\alpha\beta}\epsilon^{\mu\nu\alpha\beta})
\ee
where $\theta$ is an arbitrary constant. This term can be written as the total divergence of a vector density and hence turned into a boundary term that does not affect the classical equations of motion. It nonetheless arises naturally in the quantum formulation of the theory with important consequences. Non-zero $\theta$ leads to CP violation and contributes to the electric dipole moment of the neutron. Since the latter is experimentally very small, one has an upper bound on the value of $\theta$ of the order of $10^{-11}$. This leads to a ``fine-tuning problem'', dubbed the strong CP problem (for a review see \cite{AxionReview}).
  It provides a cautionary tale: all terms not forbidden must in principle be considered in the action, with potentially surprising results.

For the many-particle case of the standard model, requiring the global internal symmetries to be local yields
the gauge potentials and field strengths
described in the previous section to
transform as
$A_\m \rightarrow UA_\m U^\dagger + (i/g) \le( \pa_m U\ri) U^\dagger$, while
$F_{\m\n} \rightarrow U F U^\dagger$, such that $Tr(F^2)$ is invariant. Therefore the Lagrangian density of gauge fields should be of the form
$f(Tr(F^2))$, and demanding second order equations of motion narrow it own to ${\cal L} = -(1/4) Tr(F^2)$, where the negative sign ensures that
the field energy is positive, while the $1/4$ is convention such that the equations of motion with charged matter reduce to the familiar Maxwell's equations
for the $U(1)$ case.
%
%
%
In addition, one adds the Lagrangian densities for the fermions (quarks and leptons) and scalar (Higgs). Symmetry considerations require the the matter fields to be minimally coupled to the gauge fields:
\begin{eqnarray}
&&{\cal L}_{\mbox fermion}= \bar \psi \le( \gamma^\mu (i\pa_\m - A_\m) - m \ri)\psi \\
&&{\cal L}_{\mbox scalar} = -[(i\pa_\m+eA_\m)\phi^\star][(i\pa^\m - e A^\m)\phi] - m^2 \phi^\star\phi
\nonumber\\
\end{eqnarray}

Note that in the standard model of particle physics, only left handed fermions (e.g. electron, neutrino) and right handed anti-fermions (e.g. positrons, anti-neutrinos) couple to gauge fields (left and right handed fermions differ in their transformation properties under Lorentz transformations). Thus our Universe, for unknown reasons, is not left-right symmetric, showing that not all expected symmetries are realized in Nature.

The internal symmetry group of the Standard Model is composed of three sub-groups: $U(1)$, $SU(2)$ and $SU(3)$, respectively for
electromagnetism, weak and strong nuclear forces. The full symmetry group is not currently realized in nature, however, because of the presence of the Higgs boson, which develops a non-zero vacuum expectation value at low temperatures and breaks the $U(1)$ and $SU(2)$ symmetries down to just $U(1)$\footnote{There is an important technical point: the $U(1)$ symmetry in the unbroken phase of the theory is not quite the same as the $U(1)$ in the broken phase.}


%
%
%
%

%
%
%

\section{Noether's Theorem}

Noether's theorem is one of the most important, if not the most important theorem in modern physics.
It states that there exists a conserved charge corresponding to every continuous global symmetry of the
Lagrangian. A proof of this theorem goes as follows:
Consider a Lagrangian ${L}(q_a, \dot q_a, t)$, where $q_a (\dot q_a), a=1,\dots,N$ are generalized coordinates (velocities) and $t$ is the time parameter. If this Lagrangian is invariant under a symmetry transformation
$q_a \rightarrow q_a + \epsilon \delta q_a$ and $\dot q_a \rightarrow \dot q_a + \epsilon  \delta \dot q_a$, then to first order in $\epsilon$:
\bea
&& \delta L = \sum_a \left( \frac{\pa{L} }{\pa q_a} \delta q_a + \frac{\pa {L}}{\pa \dot{q}_a} \delta \dot{q}_a\right) \epsilon= 0 \nn \\
\eea
Using the Euler-Lagrange equations
\bea
\frac{\pa {L}}{\pa q_a} - \frac{d}{dt} \frac{\pa {L} }{\pa \dot q_a} = 0
\eea
changes the above to:
\bea
&& \sum_a \left(  \frac{d}{dt} \frac{\pa { L}}{\pa \dot{q}_a} \delta q_a + \frac{\pa { L}}{\pa \dot q_a} \dot q_a  \right)\epsilon = \frac{d}{dt} \left(  \sum_a \frac{\pa { L} }{\pa \dot q_a}\delta q_a \right) \epsilon
= 0 \nn \\
\eea
This implies the existence of an associate conserved charge $Q$ for each independent symmetry:
\bea
&& \sum_a \frac{\pa {L}}{\pa \dot q_a}~\delta q_a \equiv Q = \mbox{conserved}~,
\eea

Applications of Noether's theorem are ubiquitous in physics:  symmetry under spatial translations yields linear momentum as the conserved charge, while
time-translational invariance  implies energy conservation.
In the case of non-relativistic quantum mechanics, global phase invariance guarantees that probability is conserved.
For a field theory describing charged particles, the associated global gauge symmetry
yields conservation of charge. Finally, one can show that coordinate invariance and observer independence yield black hole entropy \cite{wald} as the conserved charge.
The possibilities are limited only by the number of continuous global symmetries that one is able to find.

\section{General Relativity}
\label{section:general relativity}

Einstein's theory of general relativity is a direct consequence of requiring physics to be described by second order equations that are coordinate and observer independent.  While somewhat analogous to the gauge transformations discussed above, coordinate transformations are in fact quite different mathematically and conceptually. Historically, Einstein discovered the dynamical equations governing space-time itself (Einstein equations) in 1915, just a short while before Hilbert was able to derive those equations from an action principle. The Lagrangian density is rather simple when written in terms of the curvature scalar $R \equiv g^{\m\lambda} g^{\nu\sigma} R_{\mu\nu\lambda\sigma}$, where
$R_{\m\n\lambda\sigma}= $ is the Riemann-Christoffel curvature tensor, derived from the spacetime metric
$g_{\m\n}$.
In fact it is just
\bea
{\cal L} = R+\Lambda~,
\label{Einstein action}
\eea
with $\Lambda$ being the cosmological constant, and the measure being general coordinate invariant $\int d^4x \sqrt{-g}$. (\ref{Einstein action}) is the unique invariant in 4-D that yields second order equations\cite{lovelock1969}.

{There is  another term, called the Gauss-Bonnet (GB) term, that can be added. It is higher order in the curvature but, in analogy with the $\theta$ term in Section V, it is a total divergence that does not affect the equations of motion. The Einstein equations are therefore unique in four dimensions if one demands that the equations be second order. Interestingly, this statement is special to four space-time dimensions. In higher dimensions, the Gauss-Bonnet term is not a total divergence but nonetheless yields second order equations. Moreover, the GB term is just one of an infinite class of such terms, successively higher order in curvature, that can be added in higher dimensions. The number of independent terms that yield second order equations in any given spacetime dimension $d+1$ is equal to the largest integer less than or equal to  $d/2$. So in four dimensions there is one non-trivial term,  in  five and six, there are two, namely $R$ and $GB$, whereas in seven and eight spacetime dimensions there are three, etc.} Thus, the Lovelock terms may be relevant to higher dimensional theories, such as string theory. For a recent review see \cite{LovelockReview}.

{There is another way to look at the symmetry group of general relativity. A key feature of general relativity is that it is always possible to find locally a frame in which the metric approximates the Minkowski metric.  We expect the laws of physics in such a frame to be locally the same as in flat space-time. This is the strong equivalence principle and essentially determines how matter fields couple to the geometry of space-time.
Since Minkowski space is invariant under Lorentz transformations, one should require that all the laws of physics be invariant under independent Lorentz transformations at each point in space-time. This can be formulated as a theory with a local $SO(3,1)$ gauge invariance that is completely analoguous to the gauge symmetries of the Standard Model. In order to make the geometrical fields dynamical by constructing an invariant action
one is then forced to consider gravity theories based on the Lorentz connection rather than Christoffel symbols. In the absence of torsion, the resulting theories are essentially those of the Lovelock form. A nice overview of the consequences of this formulation can be found in the review by Zanelli \cite{zanelli2005} and as well as the more recent book \cite{zanelli2016}.}

\section{Outstanding Problems}

\subsection{Beyond the Standard Model}
We currently have little or no experimental guidance as to what physics might look like beyond the strong interaction scale. In the late 20th century it was thought that at scales of about $10^{14-16}$ GeV the gauge interactions would merge into a ``Grand Unified Theory'' (GUT) based on a single symmetry group large enough to contain the symmetry group $U(1)\times SU(2)\times SU(3)$ of the Standard Model. The range of energies between the electro-weak symmetry breaking scale and the GUT scale was called the ``desert,'' because the expectation was that no new physics could emerge without disrupting the grand unification scenario. GUTs predict the decay of the proton, something that was and still is searched for but so far not found. It is not clear to what extent GUTs are currently viable. The Large Hadron Collider has provided new support for the Standard Model, but so far no guidance as to how to proceed beyond it. Moreover, the strong field limit of GR  has been recently well tested via the recent observation\cite{Ligo2015} of gravitational waves emitted during the merger of two black holes 10 billion years ago. It now appears that the desert may extend from $10^3 GeV$  all the way up to the Planck scale of $10^{19} GeV$. The result is that current attempts to go beyond the Standard Model and include gravity are primarily based on elegance, internal consistency and symmetry arguments. Although some attempts are being made to connect these theories to experiment \cite{sd},
it is a difficult thing to achieve in practice.

\subsection{Quantum Gravity}
\label{section:quantum gravity}

{Although gravity is the dominant force at macroscopic scales, it is irrelevant at the microscopic scales probed by current and forseeable  high energy experiments. However, this is expected to change at scales of the order of the Planck scale
\be
l_{pl} = \sqrt{\frac{\hbar G}{c^3}}\sim 10^{-35}~m,
\ee
Although such scales are likely to be inaccessible observationally,
one must nonetheless understand how quantum mechanics and gravity co-exist. The current lore is that gravity must be quantized. One strong motivation is that quantum mechanics is needed to cure the apparently unavoidable singularities predicted by classical theories of gravity, most notably in the early universe and at the centre of black holes. As yet no satisfactory (theoretically compelling and uniquely predictive) theory of quantum gravity exists. Standard perturbative techniques that are so successful in the Standard Model do not work. Perturbative quantum gravity is non-renormalizable: one gets infinite answers to physical questions. What's worse, these infinities can only be cured by adding more and more counter-terms to the action, leading to an infinite number of parameters that must be determined experimentally. Thus one must resort to non-perturbative techniques, which are notoriously difficult.

Significant progress along these lines has been made in the context of Loop Quantum Gravity \cite{LQGReview}, and its more recent covariant counterpart \cite{SpinFoamsReview}. LQG starts with the classical GR hamiltonian in terms of a cleverly chosen set of canonical variables and attempts to apply the standard rules of quantum mechanics.}
Several important outstanding issues remain in this approach however, including recovery of the  standard geometrical picture  in a rigorous semi-classical limit, solving the problem of dynamics  and finding a complete set of physical observables.

{The other popular and highly developed theory is string theory \cite{StringReview}, in which the point-like particles that serve as the starting point for classical dynamics are replaced by vibrating strings. This theory introduces an immense number of degrees of freedom. The good news is that there are enough of them to account for all the particles and fields we currently know and love, including gravity. The bad news is that {this embarrassment of riches} makes it  very difficult to extract unique predictions from the theory. Moreover, internal consistency of the construction requires introduction of spatial dimensions and  an extra symmetry (supersymmetry, which rotates fermions to bosons and vice-versa) that are as yet not observed. Although the theory is mathematically elegant and to some extent physically compelling, after several decades of intense research it still has a long way to go before it can be called a viable theory describing our Universe.}

\section{Conclusion: Symmetry to the Rescue?}

We have argued in this article that the requirements of
 coordinate and observer independence, gauge invariance and second order equations narrow  down our choice of theories considerably. There is not much room to deviate from those that are known and used to compute measurable quantities.
Spontaneous symmetry breaking, which gives mass to gauge bosons and fermions,
although not an unavoidable consequence of the presence of symmetries, nevertheless follows in a straightforward manner and plays an important role in our understanding of many physical systems. In the Standard Model, for example, one postulates  that nature randomly chooses one vacuum configuration among the infinite that are available for the Higgs scalar field. This is similar to the spins of a bulk ferromagnetic aligning themselves in one randomly chosen direction once the system is cooled down below its critical temperature.

We hasten to add that historically, for most cases at least, the physical laws were discovered
via widely different routes, with symmetries playing roles of varying importance.
Moreover, it is only in the case of electromagnetism and gravitation that one can claim that symmetry considerations uniquely determine the dynamics. In the context of the Standard Model one must ask of the many particle species in the standard model zoo: ``who ordered that?''

{The most important question in this context is: to what extent can symmetry guide us towards a successful theory of quantum gravity? If one looks for a theory that has the most symmetry possible, then string theory becomes an attractive choice. Since it can be formulated in terms of a conformally invariant theory in two spacetime dimensions, string theory comes endowed with an infinite dimensional internal symmetry group. In addition, the strings live in a higher dimensional ``target space'' that manifests not only the higher dimensional coordinate transformation group as a symmetry, but also supersymmetry. However, unless the existence of these symmetries can be verified experimentally, string theory may provide an example of a theory that has more symmetry than is ultimately useful. To adapt another quote due to Einstein: ``a theory should be as symmetric as possible, but not more symmetric''\footnote{The actual quote commonly ascribed to Einstein is something like: "Everything should be made as simple as possible, but no simpler" .}.

As an alternative, and certainly less ambitious starting point, one can try to use symmetry considerations to determine which classical Lagrangian density or equation of motion to quantize.
History has shown, for example, that while the Fermi theory of weak interactions and the Salam-Weinberg model both have the
required space-time symmetries and work well, only the second is
 renormalizable \cite{SM}. Similarly, the Raychaudhuri equation, which also has all the above symmetries and is a first order equation in the geodesic expansion $\theta$, is completely equivalent to the Einstein equation if one assumes in addition the correctness of the Bekenstein-Hawking area law \cite{jacobson}. Thus it may be possible that the former, when quantized, is renormalizable, and furnishes meaningful, measurable quantities in the language of the latter.

To summarize, symmetry considerations have taken us a long way towards understanding what choices exist for describing the Universe and will continue to do so for the foreseeable future. Whether or not they will ultimately lead us to a unique theory remains to be seen. It is certainly a question worthy of active consideration.

%

\vs{.2cm}
\no {\bf Acknowledgments}

\no
This work is supported by the Natural Sciences and Engineering
Research Council of Canada.
We thank R. Danos, A. Frey, J. Zanelli  and J. Ziprick for comments that helped improve the manuscript.
We thank our past and present collaborators for many fruitful discussions.



\end{document}